\documentclass[a4paper]{jpconf}
\usepackage{amsfonts,amssymb,amsmath}
\usepackage{graphicx}

\newcommand{\be}{\begin{equation}}
\newcommand{\bea}{\begin{eqnarray}}
\newcommand{\ee}{\end{equation}}
\newcommand{\eea}{\end{eqnarray}}
\newcommand{\sla}{\slash \hspace{-0.22cm}}

\def\s#1{{\scriptscriptstyle #1}}

\def\1eq#1{Eq.~(\ref{#1})}
\def\2eqs#1#2{Eqs.~(\ref{#1}) and~(\ref{#2})}
\def\3eqs#1#2#3{Eqs.~(\ref{#1}), (\ref{#2}) and~(\ref{#3})}
\def\4eqs#1#2#3#4{Eqs.~(\ref{#1}), (\ref{#2}), (\ref{#3}) and~(\ref{#4})}

\def\fig#1{Fig.~\ref{#1}}

\def\ie{{\it i.e.}, }
\def\eg{{\it e.g.}, }

\def\Vbqq{\widetilde{V}}

\def\NV{\gfullb'}

\def\NP{\widetilde{V}}

\def\gfullb{\widetilde{\Gamma}}

\begin{document}
\title{Unraveling the organization of the QCD tapestry}

\author{J. Papavassiliou}

\address{Department of Theoretical Physics and IFIC, 
University of Valencia and CSIC,\\
E-46100, Valencia, Spain}

\ead{Joannis.Papavassiliou@uv.es}

\begin{abstract}
I review some key aspects of the ongoing progress in our understanding of the infrared dynamics of the QCD Green's 
functions, derived from the close synergy between Schwinger-Dyson equations 
and lattice simulations. 
Particular attention is dedicated to the elaborate nonperturbative mechanisms that endow the 
fundamental degrees of freedom (quarks and gluons) with dynamical masses.  
In addition, 
the recently established connection between the effective interaction  
obtained from the gauge sector of the theory and that needed for the veracious description 
of the ground-state properties of hadrons is briefly presented.

\end{abstract}

\section{Introduction}

\bigskip
\indent

Sixty years after their invention~\cite{Yang:1954ek}, 
Yang-Mills theories occupy the center stage of 
elementary particle physics, providing  a fundamental description for 
an impressive array of physical phenomena. 
The remarkable property of asymptotic freedom makes these theories particularly 
attractive, and, in a sense, self-contained. However, while perturbation theory works well 
in their ultraviolet regime, the infrared dynamics 
represent a challenge of notorious complexity. 
In fact, in the case of QCD~\cite{Marciano:1977su},
the most characteristic phenomena, such as 
confinement, mass generation, and bound-state formation, are purely nonperturbative.

In this presentation we will focus on the paradigm-shifting picture that emerges 
for some of the aforementioned phenomena 
from the systematic study of Green's functions within a global framework,  
where the analysis of the Schwinger-Dyson equations (SDEs) is complemented 
and refined by inputs obtained from large-volume 
lattice simulations.

The present study is restricted to the special case of the Landau gauge, where 
the vast majority of recent lattice simulations have been performed.
The SDE part of the problem is addressed within the formalism obtained  from the 
fusion between the pinch technique (PT)~\cite{Cornwall:1981zr,Cornwall:1989gv,Pilaftsis:1996fh,Binosi:2002ft,Binosi:2009qm} 
and the background field method (BFM)~\cite{Abbott:1980hw}, 
denoted simply as ``PT-BFM''
(for a representative sample of different approaches, 
see~\cite{Alkofer:2000wg,Aguilar:2004sw,Fischer:2006ub,Braun:2007bx,Boucaud:2008ji,Boucaud:2008ky,Dudal:2008sp,Fischer:2008uz,Pennington:2011xs,Serreau:2012cg,Tissier:2011ey,Siringo:2014lva}).

This presentation is organized as follows. 
In section~\ref{QBsec} we review the PT-BFM formalism, and highlight its  
most salient features. Then, 
in section~\ref{remind} we summarize the subtle filed-theoretic 
mechanism that leads to the gauge-invariant generation of a dynamical gluon mass. 
In section~\ref{ghost} we present a brief overview of the 
impressive coincidence achieved between SDE calculations and lattice simulations of the ghost propagator,
while in section~\ref{quark} we address some of the intricate issues surrounding the 
dynamical generation of the constituent quark masses.
Then, in section~\ref{hadrons} we review a recent {\it ab-initio} derivation of 
the interaction kernel that has been extensively used in the contemporary 
description of hadron phenomenology. 
Finally, in section~\ref{concl} we summarize our conclusions.

\section{\label{QBsec} The PT-BFM formalism: a powerful framework for studying SDEs}

\bigskip
\indent

In what follows 
we will work exclusively in the Landau gauge, where the gluon propagator 
assumes the totally transverse form 
\be
i\Delta_{\mu\nu}(q)=- i P_{\mu\nu}(q)\Delta(q^2);\qquad
P_{\mu\nu}(q)=g_{\mu\nu}- q_\mu q_\nu/q^2 \,.
\label{prop}
\ee 
The scalar  form factor
$\Delta(q^2)$  is  related to  the  all-order  gluon
self-energy $\Pi_{\mu\nu}(q)$. Specifically, as a consequence of the BRST symmetry, 
$\Pi_{\mu\nu}(q)$  
is transverse, both perturbatively and nonperturbatively; one has then   
\be
q^\mu\Pi_{\mu\nu}(q)=0;\qquad \Pi_{\mu\nu}(q)=\Pi(q^2)P_{\mu\nu}(q),
\label{trangen}
\ee
and $\Delta^{-1}(q^2) = q^2 + \Pi(q^2)$.
In addition, for later convenience, we introduce the {\it inverse} 
of the gluon dressing function, to be denoted by  $J(q^2)$, namely
\be
\Delta^{-1}({q^2})=q^2 J(q^2).
\label{defJ}
\ee

\begin{figure}[!t]
\begin{center}
\includegraphics[scale=1.0]{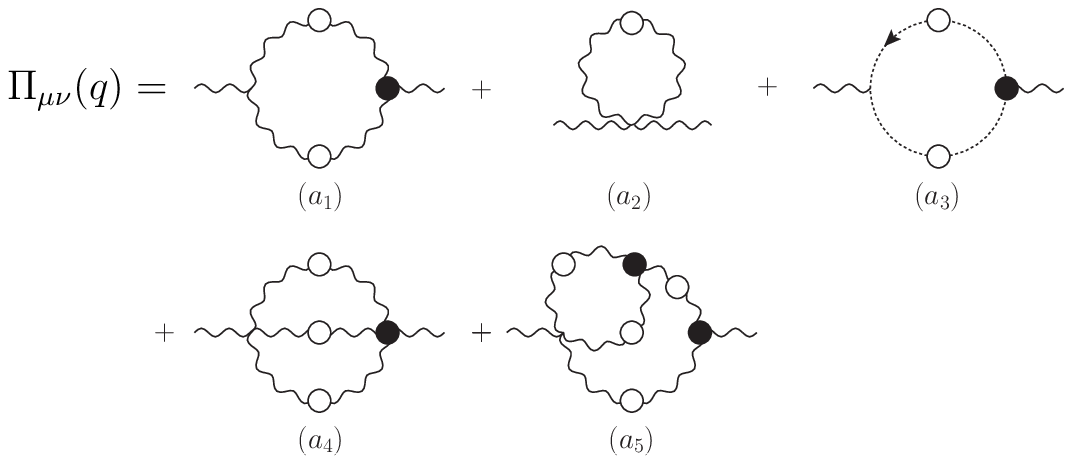}
\caption{\label{glSDEQQ} The conventional SDE of the standard gluon propagator ($QQ$). 
Black blobs represent fully dressed one-particle irreducible  vertices, 
whereas the white ones denote fully dressed propagators.}
\end{center}
\end{figure}

The nonperturbative dynamics of the gluon propagator are governed by the corresponding SDE. In particular, 
within the conventional formulation~\cite{Roberts:1994dr,Maris:2003vk}, the gluon self-energy $\Pi_{\mu\nu}(q)$ 
is given by the fully dressed diagrams shown in \fig{glSDEQQ}.
One longstanding difficulty with this equation is that it cannot be truncated in any obvious way without 
compromising the validity of  \1eq{trangen}. This happens because the conventional 
fully dressed vertices appearing in the diagrams of \fig{glSDEQQ} satisfy complicated Slavnov-Taylor identities
(STIs), and it is only after the inclusion of all diagrams that \1eq{trangen} may be enforced.

\begin{figure}[!t]
\begin{center}
\includegraphics[scale=1.0]{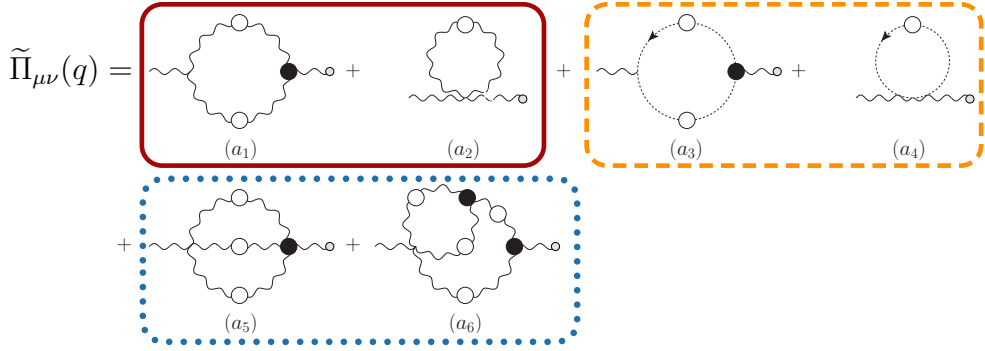}
\caption{\label{glSDE} The SDE obeyed by the $QB$ gluon propagator. Black blobs represents fully dressed 1-PI vertices; 
the small gray circles appearing on the external legs (entering from the right, only!) 
are used to indicate background gluons. The diagrams contained in each box form  
individually transverse subsets.}
\end{center}
\end{figure}

Instead,
the formulation of this SDE in the context of the  PT-BFM formalism furnishes 
considerable advantages, because  it allows for a systematic truncation 
that respects manifestly, and at every step, the crucial identity of \1eq{trangen}.
The deeper reason why such a truncation becomes possible may be traced back  
to the fact that the PT-BFM Green's functions satisfy (by construction) 
Abelian-like Ward identities (WIs) instead of STIs. In particular, if we employ the BFM language and 
distinguish the gluon fields into background ($B$) and quantum ($Q$) ones, vertices 
such as $BQQ$, $B{\bar c}c$, and $BQQQ$, to be denoted by 
 $\widetilde{\Gamma}_{\alpha\mu\nu}(q,r,p)$ ,   $\widetilde{\Gamma}_\alpha(q,r,-p)$, 
and $\widetilde{\Gamma}^{mnrs}_{\mu\nu\rho\sigma}(q,r,p,t)$, respectively, satisfy the simple WIs~\cite{Aguilar:2006gr} 
\be \label{WIBQ2}
q^\alpha \widetilde{\Gamma}_{\alpha\mu\nu}(q,r,p) = i\Delta_{\mu\nu}^{-1}(r) - i\Delta_{\mu\nu}^{-1}(p), 
\ee
\be \label{WIBcc}
q^\alpha \widetilde{\Gamma}_\alpha(q,r,-p) = D^{-1}(q+r) - D^{-1}(r),
\ee
and 
\bea \label{WIBQ3}
q^\mu \widetilde{\Gamma}^{mnrs}_{\mu\nu\rho\sigma}(q,r,p,t) &=& f^{mse}f^{ern} \Gamma_{\nu\rho\sigma}(r,p,q+t) + f^{mne}f^{esr}\Gamma_{\rho\sigma\nu}(p,t,q+r) \nonumber \\
&+& f^{mre}f^{ens} \Gamma_{\sigma\nu\rho}(t,r,q+p).
\eea
These particular vertices (instead of the conventional ones) appear in the SDE of \fig{glSDE}, which 
controls the self-energy $\widetilde\Pi_{\mu\nu}(q)$ of the 
the mixed  background-quantum gluon propagator ($BQ$), 
denoted by  $\widetilde\Delta(q^2)$. Consequently, it is relatively straightforward to prove  
the block-wise transversality of $\widetilde\Pi_{\mu\nu}(q)$, namely~\cite{Aguilar:2006gr}  
\be
q^{\mu} [(a_1) + (a_2)]_{\mu\nu} = 0;\qquad
q^{\mu} [(a_3) + (a_4)]_{\mu\nu} = 0; \qquad
q^{\mu} [(a_5) + (a_6)]_{\mu\nu} = 0.
\label{boxtr2}
\ee 

From~\1eq{glSDE} it is clear that 
the SDE of $\widetilde\Delta(q^2)$ contains inside its defining diagrams 
the full propagator $\Delta(q^2)$; therefore, in that sense, it cannot be 
considered as a dynamical equation for neither  $\widetilde\Delta(q^2)$ nor $\Delta(q^2)$.
At this point a crucial identity relating $\Delta(q^2)$ and   $\widetilde{\Delta}(q^2)$~\cite{Grassi:1999tp, Binosi:2002ez} 
enters into the game.  
Specifically, one has that  
\be
\Delta(q^2) = [1 + G(q^2)] \widetilde{\Delta}(q^2), 
\label{BQIs}
\ee
where $G(q^2)$ is the $g_{\mu\nu}$ co-factor of a certain two-point function that  
originates from the ghost sector of the theory.

The fundamental observation put forth in a series of works~\cite{Aguilar:2006gr,Binosi:2007pi,Binosi:2008qk}
is that one may use the SDE for $\widetilde{\Delta}(q^2)$, 
take advantage of its improved truncation properties,  and then 
convert it to an equivalent equation for $\Delta(q^2)$ (the propagator simulated on the lattice) 
by means of the special relation given in \1eq{BQIs}. 
Thus, the PT-BFM version of the SDE for the conventional gluon propagator reads  
\be
\Delta^{-1}(q^2){ P}_{\mu\nu}(q) = 
\frac{q^2 {P}_{\mu\nu}(q) + i\,\sum_{i=1}^{6}(a_i)_{\mu\nu}}{1+G(q^2)}.
\label{sde}
\ee

\section{\label{remind} Dynamical gluon mass generation}

\bigskip
\indent

The possibility that Yang-Mills theories generate dynamically a gluon mass was first proposed 
and explored in the early eighties~\cite{Cornwall:1981zr,Bernard:1982my,Donoghue:1983fy}, 
and has attracted particular attention in recent years, 
mainly due an accumulation of hard evidence obtained from large-volume  
lattice simulations, both in SU(3)~\cite{Bogolubsky:2009dc,Bogolubsky:2007ud,Bowman:2007du,Oliveira:2009eh}
and in SU(2)~\cite{Cucchieri:2007md,Cucchieri:2007rg,Cucchieri:2009zt,Cucchieri:2010xr}
(for a different lattice approach, see~\cite{Philipsen:2001ip}).

\begin{figure}[!h]
\begin{minipage}[b]{0.45\linewidth}
\centering
\includegraphics[scale=0.35]{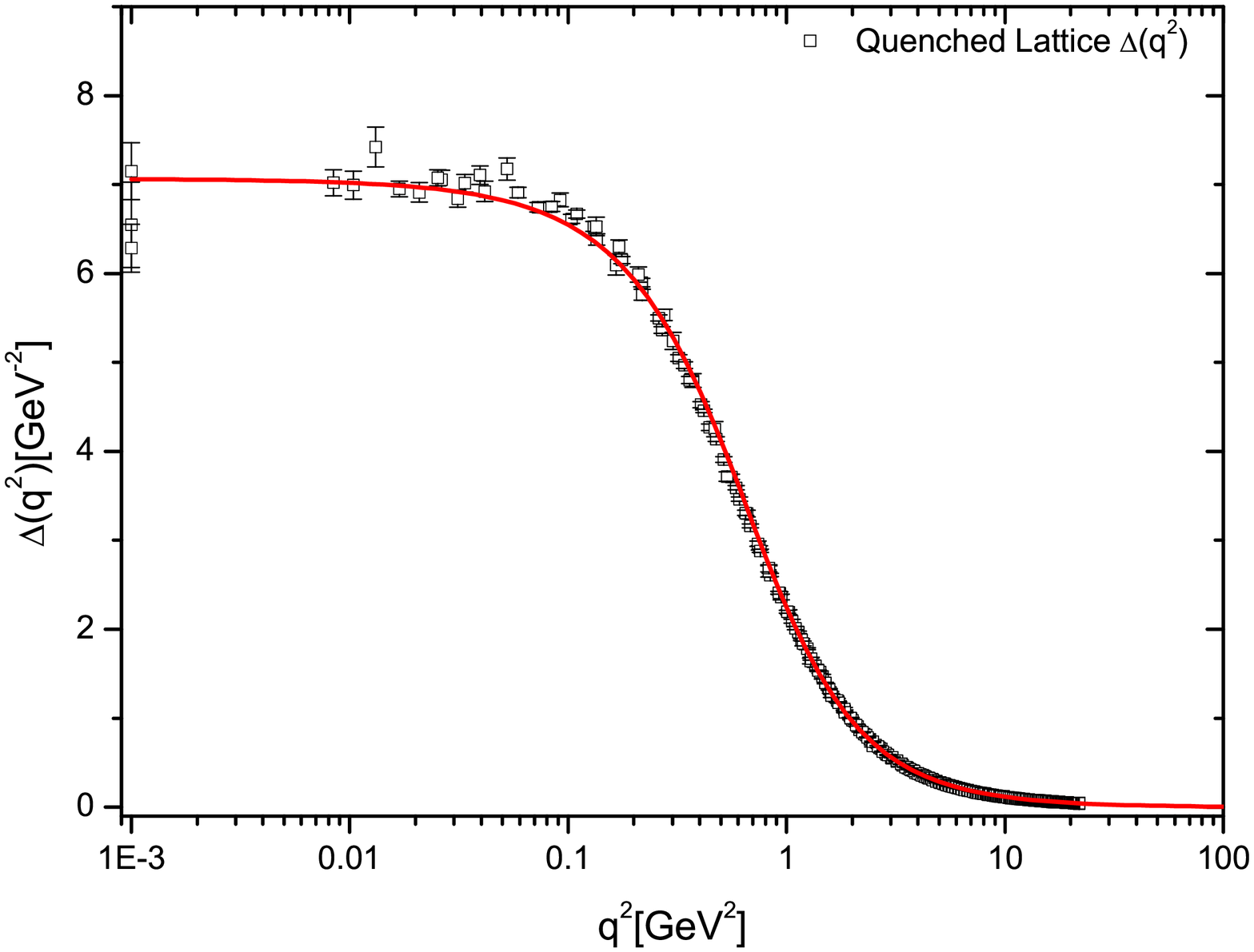}
\end{minipage}
\hspace{0.5cm}
\begin{minipage}[b]{0.50\linewidth}
\includegraphics[scale=0.35]{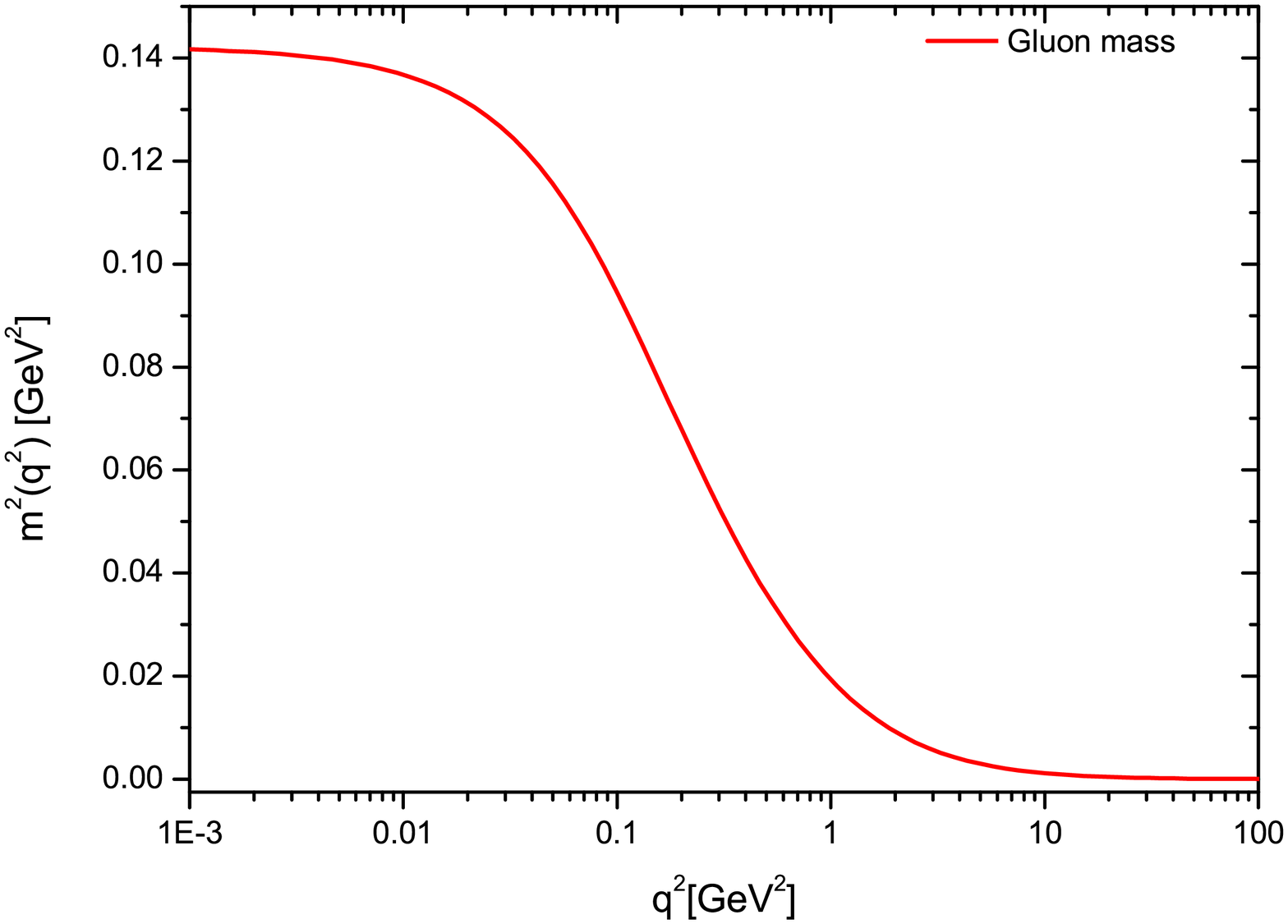}
\end{minipage}
\vspace{-0.25cm}
\caption{\label{fig:gluon_mass} The  quenched lattice gluon propagator $\Delta(q^2)$  (left panel), and the corresponding gluon mass (right panel).}
\end{figure}

What these simulations reveal is that the Landau gauge
gluon propagator saturates in the deep infrared at a fixed nonvanishing value, as shown on the left panel of~\fig{fig:gluon_mass},
which is the smoking gun signal for gluon mass generation. A conclusive demonstration of how this may happen 
at the level of the gluon SDE within the PT-BFM framework has been given in~\cite{Aguilar:2008xm}.

From the kinematic point of view 
we will describe the transition 
from a massless to a massive gluon propagator by implementing the replacement  
(in Euclidean space)
\be
\Delta^{-1}(q^2) = q^2 J(q^2) \quad \longrightarrow\quad  \Delta^{-1}(q^2)=q^2 J(q^2)+m^2(q^2),
\label{massive}
\ee
where $m^2(q^2)$ is the (momentum-dependent) dynamically generated mass, 
with the property that $m^2(0) >0$; evidently, $\Delta^{-1}(0) =m^2(0)$. 

The actual field-theoretic mechanism that permits the generation of the term $m^2(q^2)$ 
can be traced back to the seminal work of Schwinger~\cite{Schwinger:1962tn,Schwinger:1962tp}, 
which may be summarized 
in the statement that the vacuum polarization of  
a gauge boson that is massless
at the level of the original Lagrangian may develop a massless pole, whose 
residue will be eventually identified with $m^2(0)$. 
In the case of Yang-Mills theories,
the origin of the aforementioned poles is 
due to purely nonperturbative dynamics: for sufficiently strong binding, 
the mass of certain (colored) bound states 
may be reduced to zero~\cite{Jackiw:1973tr,Jackiw:1973ha,Cornwall:1973ts,Eichten:1974et,Poggio:1974qs}.  
As has been shown in~\cite{Aguilar:2011xe} through a detailed study of the Bethe-Salpeter equation 
that controls the formation of such a composite massless pole,  
this dynamical scenario is indeed realized.

\begin{figure}[t]
\center{\includegraphics[scale=0.6]{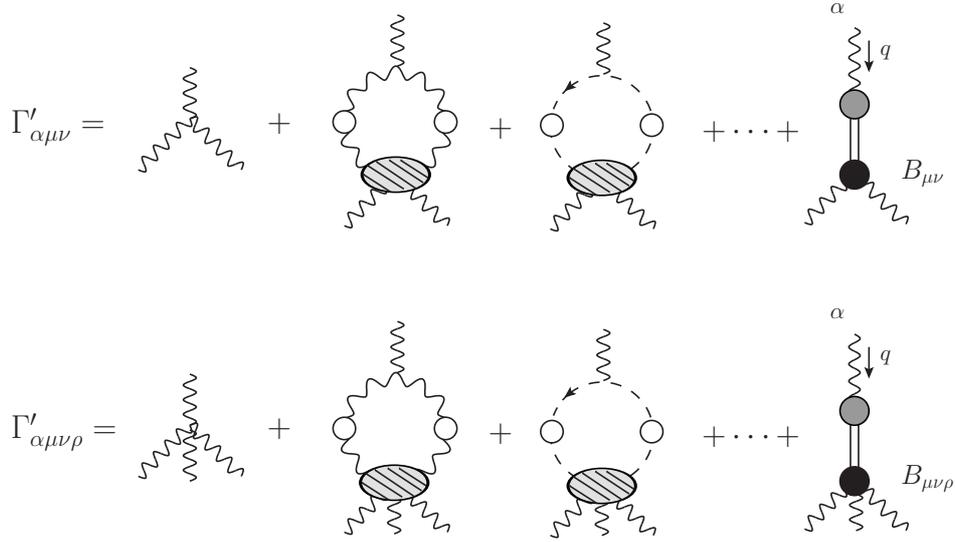}}
\caption{The SDEs for the full three and four gluon vertices in
 the presence of their corresponding pole parts.}
\label{skeletonexpansions}
\end{figure}

In addition to  triggering the Schwinger mechanism, these bound-state poles 
act as composite, longitudinally coupled Nambu-Goldstone bosons, maintaining gauge invariance; 
notice, however, that they differ from ordinary Nambu-Goldstone bosons 
as far as their origin is concerned, since they are not associated 
with the spontaneous breaking of any continuous symmetry. In particular, 
gauge invariance requires that the replacement described schematically in \1eq{massive} 
be accompanied by a simultaneous replacement of all relevant vertices by 
\be
\gfullb \quad \longrightarrow\quad   \NV = \gfullb + \Vbqq ,  
\label{nv}
\ee
where $\Vbqq$ contains the aforementioned massless poles (see Fig.~\ref{skeletonexpansions}), and guarantees 
that the new vertex $\NV$ 
satisfies the same WIs as $\gfullb$, but now replacing the 
gluon propagators appearing on their rhs by massive ones. 
For example, in order to maintain \1eq{WIBQ2} intact, the corresponding $\Vbqq$
must satisfy the WI
\be
q^\alpha \NP_{\alpha\mu\nu}(q,r,p)= m^2(r^2)P_{\mu\nu}(r) - m^2(p^2)P_{\mu\nu}(p);
\label{winp}
\ee
similarly, when contracted with respect to the other two momenta, $\NP_{\alpha\mu\nu}(q,r,p)$ 
satisfies the STIs necessary to enforce the validity of the corresponding STIs 
satisfied by $\gfullb_{\alpha\mu\nu}(q,r,p)$, but now with massive instead of massless 
gluon propagators on the rhs.

The fact that the massless poles must be longitudinally coupled is reflected at the level of 
$\NV$ through the condition 
 \be
P^{\alpha'\alpha}(q) P^{\mu'\mu}(r) P^{\nu'\nu}(p) \Vbqq_{\alpha'\mu'\nu'}(q,r,p)  = 0,
\label{totlon}
\ee   
and a exactly analogous relation for the four-gluon  
$\Vbqq_{\alpha\mu\nu\rho}(q,r,p,t)$. These latter conditions guarantee 
that the massless poles decouple completely from the on-shell $S$-matrix or  
other physical observables. Moreover, when \1eq{totlon} is combined 
with \1eq{winp} and the additional STIs not reported here, one obtains the 
exact closed expression for $\Vbqq_{\alpha'\mu'\nu'}(q,r,p)$~\cite{Ibanez:2012zk}.

The inclusion of the aforementioned vertices into the gluon SDE 
leads to its separation into two integral equations of the generic form
\bea
J_m(q^2) &=& 1+ \int_{k} {\cal K}_1 (q^{2},m^2,\Delta_m),
\nonumber\\
m^{2}(q^2) &=&  \int_{k} {\cal K}_2 (q^{2},m^2,\Delta_m),
\label{separ}
\eea
such that $q^{2} {\cal K}_1 (q^{2},m^2,\Delta_m) \to 0$, as $q^{2}\to 0$, 
whereas ${\cal K}_2 (q^{2},m^2,\Delta_m)\neq 0$ in the same limit,
precisely because it includes the  $1/q^2$ terms contained inside the $\widetilde{V}$ terms.  
This set of equations bares a close analogy to those studied in the case of the 
more familiar phenomenon of quark mass generation; in particular, $m^{2}(q^2)$ plays the role of 
the quark mass function (denoted by $B(p^2)$ in~\1eq{qpropAB}).

The explicit closed form of the integral equation that determines $m^{2}(q^2)$, 
together with a variety of related theoretical considerations,  
have been presented in a series of works~\cite{Aguilar:2009ke,Binosi:2012sj,Aguilar:2014tka}. The upshot is that one obtains 
a dynamical gluon mass of the form shown on the right panel of~\fig{fig:gluon_mass}, whose functional 
dependence may be fitted by 
\be
m^2(q^2) = \frac{m_0^2}{1+ (q^2/{\mathcal M}^2)^{1+p}}\,,
\label{fit_mass}
\ee
with $m_0=375\,\mbox{MeV}$,  $p=0.11$ and ${\mathcal M}=431\, \mbox{MeV}$. 

The nontrivial momentum dependence of the gluon mass is mainly responsible  
for the fact that, contrary to a propagator with a constant mass, the gluon propagator 
of~\fig{fig:gluon_mass} displays an inflection point. It turns out that the presence of such  
a feature is a sufficient condition for the spectral density of the gluon propagator, $\rho$, to be 
non-positive definite.  

\begin{figure}[b]
\begin{minipage}[b]{0.45\linewidth}
\centering
\includegraphics[scale=0.34]{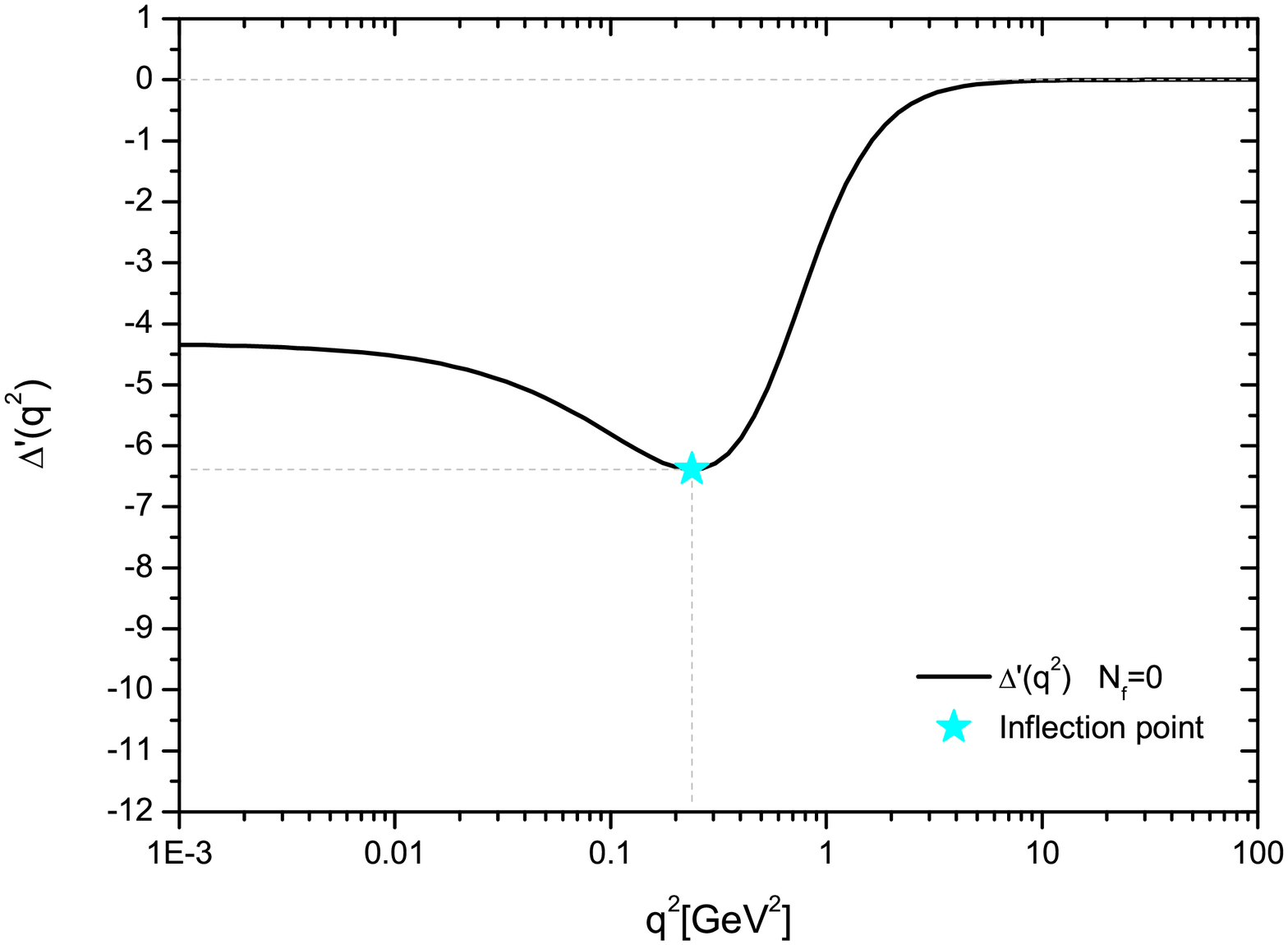}
\end{minipage}
\hspace{0.5cm}
\begin{minipage}[b]{0.50\linewidth}
\includegraphics[scale=0.34]{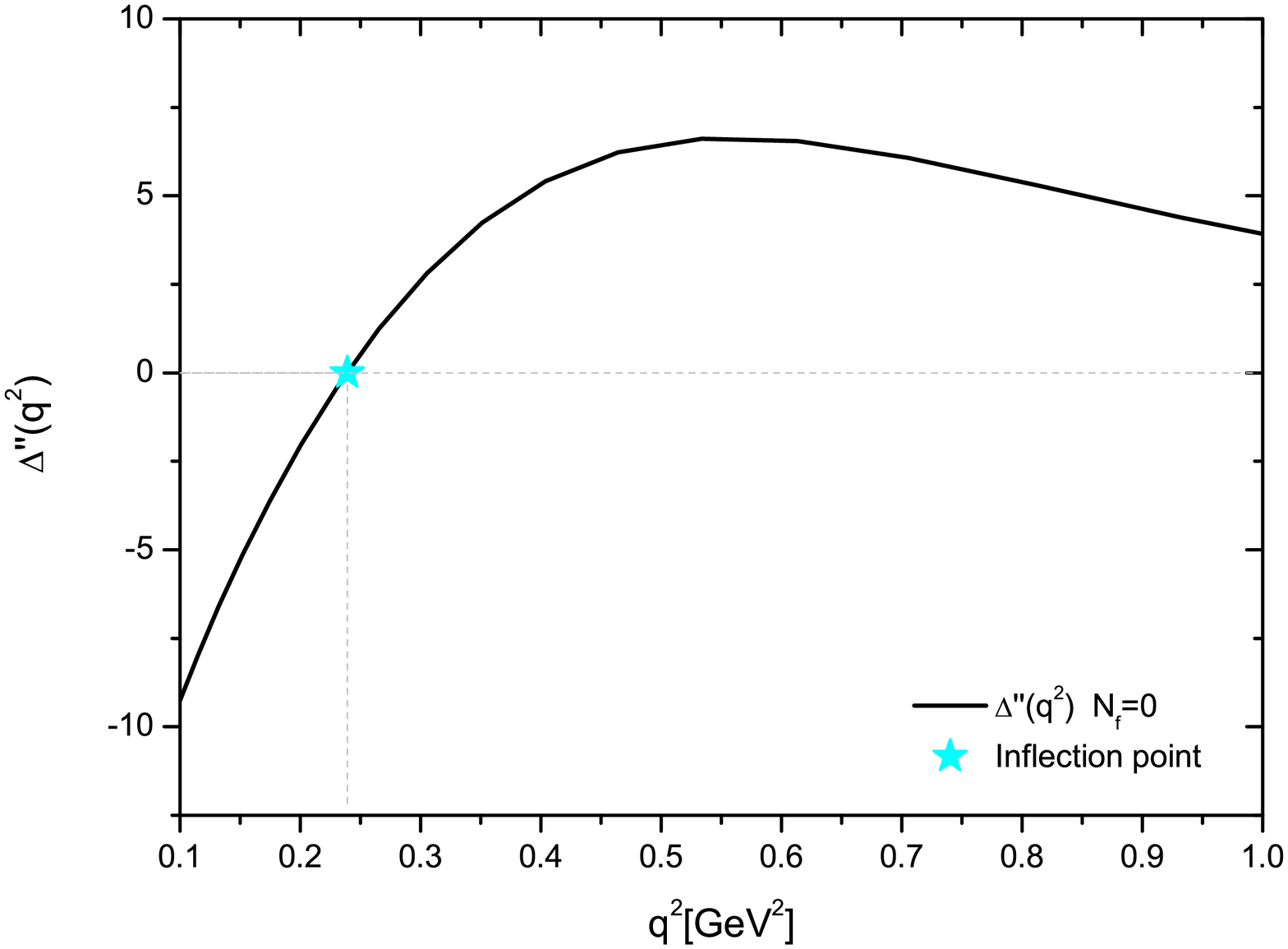}
\end{minipage}
\vspace{-0.25cm}
\caption{\label{fig:derivatives} The first and second derivatives of the gluon propagator.}
\end{figure}

Specifically, the K\"all\'en-Lehman representation of the gluon propagator reads  
\be
\Delta(q^2)= \int^{\infty}_{0} d\sigma \frac{\rho(\sigma)}{q^2+ \sigma} \,,
\label{spectral}
\ee
and if $\Delta(q^2)$  has an inflection point at $q^2_{\star}$, then its second 
derivative vanishes at that point (see~\fig{fig:derivatives}), namely~\cite{footnote}
\be
\Delta^{\prime\prime}(q^2_{\star})= 2\int^{\infty}_{0} d\sigma \frac{\rho(\sigma)}{(q^2_{\star}+ \sigma)^3} = 0 \,.
\label{spectral2}
\ee
Given that $q^2_{\star}>0$, then $\rho(\sigma)$ is forced to reverse sign at least once. The non-positivity of  $\rho(\sigma)$, in turn, may be 
interpreted as a signal of confinement (see~\cite{Cloet:2013jya}, and references therein), since the breeching of the axiom of reflection positivity excludes the gluon 
from the Hilbert space of observables states (for related works, see~\cite{Del Debbio:1996mh,Szczepaniak:2001rg,Langfeld:2001cz,Szczepaniak:2003ve,Greensite:2003bk,Gattnar:2004bf,Greensite:2011pj}).

As can be seen in~\fig{fig:derivatives}, the first derivative of $\Delta(q^2)$ displays a minimum at $q^2_{\star} = 0.238 \, \mbox{GeV}^2$, 
and, consequently, the second derivative vanishes at that same point.

\section{\label{ghost}Ghost sector}

\bigskip
\indent

The nonperturbative behavior of the ghost propagator, $D(p^2)$, has been the focal point of 
intensive studies, because in the past it had been linked to a certain confinement scenario~\cite{Alkofer:2000wg},
whose realization required that the corresponding 
ghost dressing function, $F(p^2) = p^2 D(p^2)$, should 
diverge at $p^2=0$. However,
a plethora of high quality lattice 
simulations (for SU(3)~\cite{Bogolubsky:2007ud,Bowman:2007du,Bogolubsky:2009dc,Oliveira:2009eh}
and for SU(2)~\cite{Cucchieri:2007md,Cucchieri:2007rg,Cucchieri:2009zt,Cucchieri:2010xr})
have conclusively established that $F(p^2)$ is finite 
at the origin. In fact, the finiteness of $F(p^2)$ is tightly interwoven with the  
massiveness of the gluon, because it is precisely the presence of the 
gluon mass that tames the logarithm 
associated with $F(p^2)$, and prevents it from diverging in the infrared~\cite{Aguilar:2008xm}.  

\begin{figure}[!h]
\begin{center}
\includegraphics[scale=0.65]{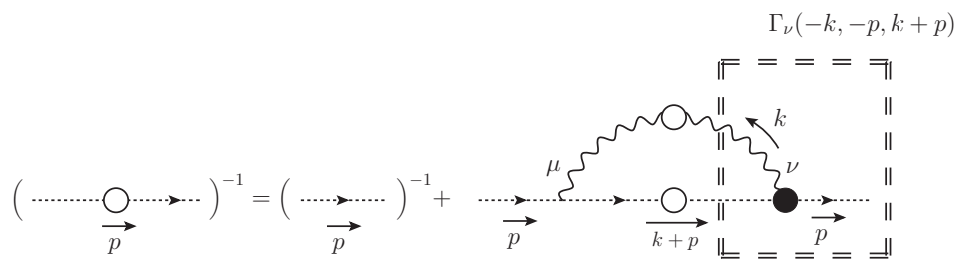}
\end{center}
\caption{\label{fig:ghostsde} The SDE for the ghost propagator given by~\1eq{SDgh}. The white blobs represent the
fully dressed gluon and ghost propagators, while the black blob denotes the dressed ghost-gluon
vertex.}
\end{figure}
Even though the finiteness of $F(p^2)$ 
may be explained qualitatively in the above relatively simple terms, a more strenuous effort   
is required in order to obtain 
from the corresponding SDE the full curve of $F(p^2)$ observed on the lattice.

\begin{figure}[t]
\begin{minipage}[b]{0.45\linewidth}
\centering
\hspace{-0.5cm}
\includegraphics[scale=0.37]{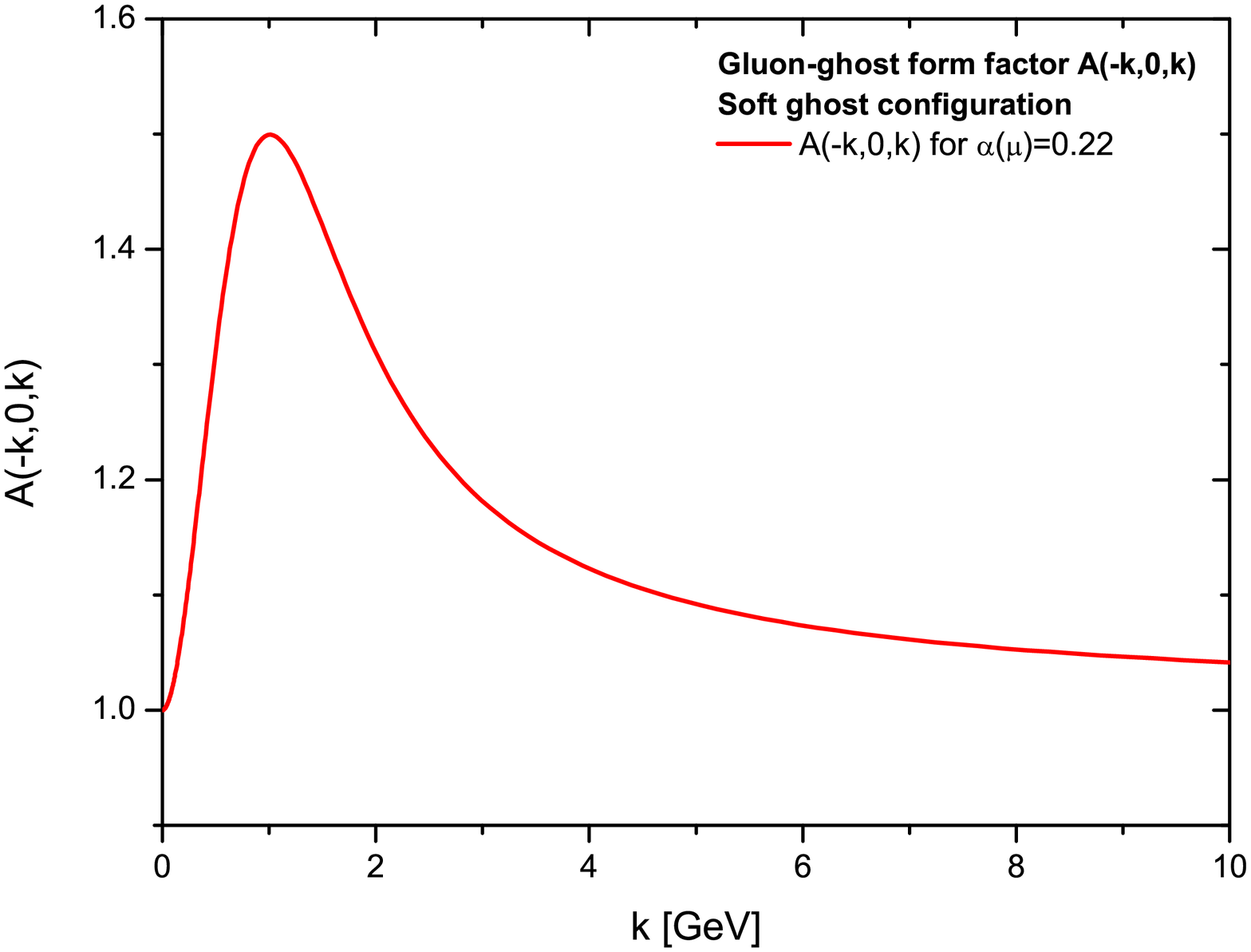}
\end{minipage}
\hspace{0.5cm}
\begin{minipage}[b]{0.50\linewidth}
\hspace{0.5cm}
\includegraphics[scale=0.37]{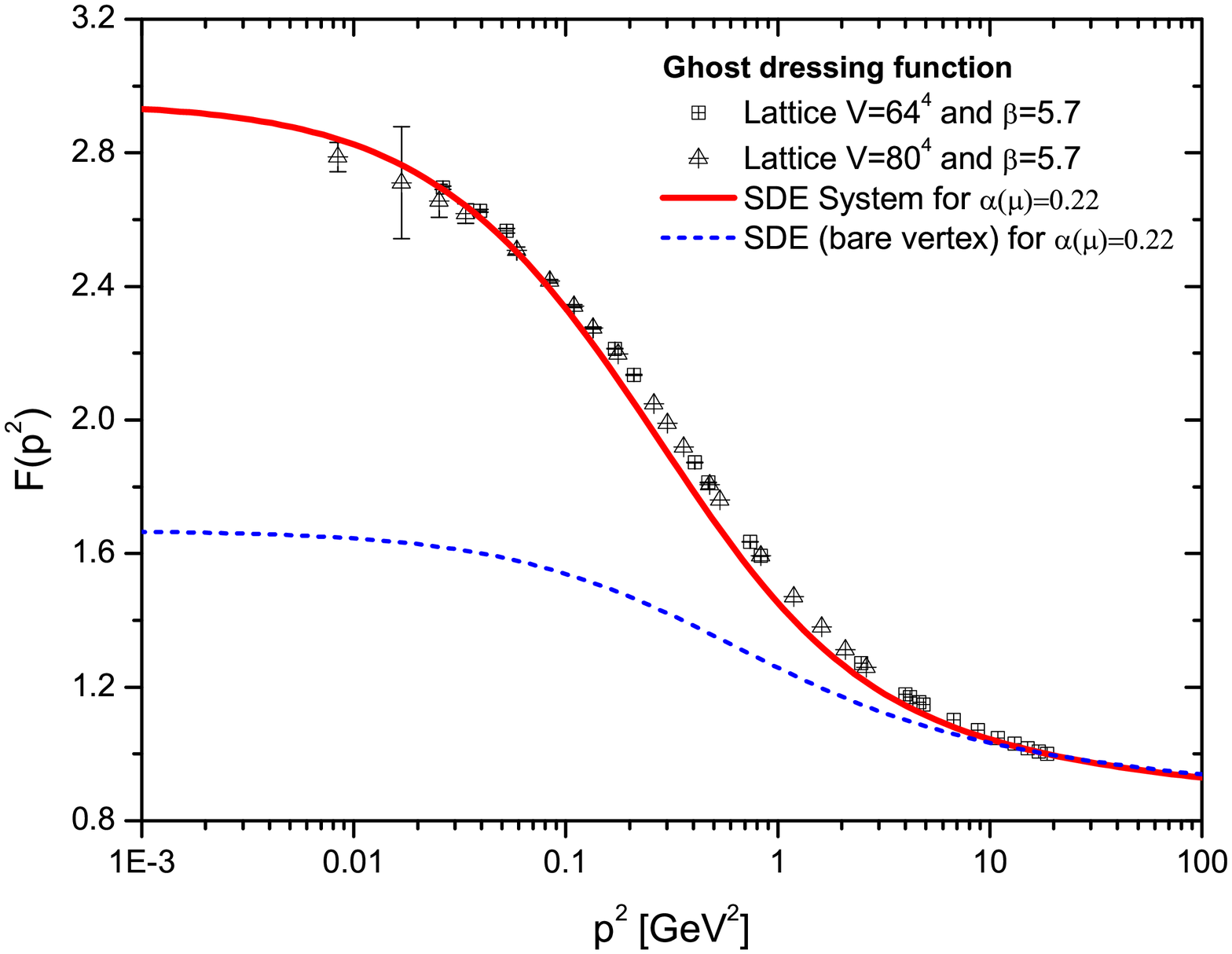}
\end{minipage}
\vspace{-0.25cm}
\caption{ 
{\it Left panel}: The numerical result for the form factor $A(-k,0,k)$ in the soft ghost configuration (red continuous line)~\cite{Aguilar:2013xqa}.  
{\it Right panel}: The $F(p^2)$ obtained from 
the ghost SDE with a bare ghost-gluon vertex (blue dashed curve) and the corresponding solution when 
 the $A(-k,0,k)$ is used (red continuous line) , using in both cases $\mu = 4.3$ GeV and $\alpha_s(\mu)=0.22$. 
The lattice data are from Ref.~\cite{Bogolubsky:2007ud}. }
\label{fig3}
\end{figure}

The SDE for the ghost propagator is diagrammatically represented in the~\fig{fig:ghostsde},
and assumes the form
\be
iD^{-1}(p^2) = ip^2 - g^2 C_{\rm {A}}  \int_k
\Gamma_{\mu}^{[0]}(k,-k-p,p)\Delta^{\mu\nu}(k)\Gamma_{\nu}(-k,-p,k+p) D(k+p)\,.
\label{SDgh}
\ee
In the above equation  $C_A$ represents the Casimir eigenvalue of the adjoint representation ($N$ for $SU(N)$), 
$d=4-\epsilon$ is the space-time dimension, and we have introduced the integral measure
\be
\int_{k}\equiv\frac{\mu^{\epsilon}}{(2\pi)^{d}}\!\int\!\mathrm{d}^d k,
\label{dqd}
\ee
with $\mu$ the 't Hooft mass. 
The vertex $\Gamma_{\nu}(-k,-p,k+p)$ is the fully dressed ghost-gluon vertex, whose 
tensorial decomposition is given by ($r=k+p$)
\be
{\Gamma}_{\nu}(-k,-p,r) =  A(-k,-p,r)\, p_{\nu} + B(-k,-p,r) \,k_{\nu} \,;
\label{Gtens}
\ee 
at tree-level, the two form factors assume the values 
\mbox{$A^{[0]}(-k,-p,r)=1$} and $B^{[0]}(-k,-p,r)=0$,  reproducing the standard bare vertex  
$\Gamma^{[0]}_{\nu}=p_\nu$.
Clearly, due to the full transversality of $\Delta_{\mu\nu}(k)$, 
any reference to the form factor $B$  
disappears from the ghost SDE of \1eq{SDgh}, and one obtains.
\be
F^{-1}(p^2) = 1 +ig^2 C_{\rm {A}} \int_k\, \left[1-\frac{(k\cdot p)^2}{k^2p^2}\right] A(-k,-p,k+p)\Delta (k)  D(k+p) \,.
\label{tt2}
\ee
The dependence of this equation on both $\Delta (k)$ and $A(-k,-p,k+p)$ is just another example of the perennial difficulty 
inherent to the SDE framework, namely the fact that, in principle, all Green's functions
of the theory are coupled to each other. In the case at hand, $\Delta (k)$ is known from the lattice (\fig{fig:gluon_mass}),
and there is no need to resort to its own SDE. However, $A$ is more problematic, since there are no lattice data 
for the kinematic configuration relevant to this problem (for SU(2), see~\cite{Cucchieri:2004sq,Cucchieri:2006tf,Cucchieri:2008qm}). 
The obvious easy way out, namely  
the use of the tree-level value for $A$ is quantitatively 
insufficient: one obtains the result shown on the right panel of~\fig{fig3} (blue dashed curve), 
which, even though qualitatively correct, it is clearly considerably suppressed compared to the lattice data. 

To ameliorate this situation, an approximate version of the SDE equation that 
controls the evolution of $A$ has been considered in~\cite{Aguilar:2013xqa},
in the special kinematic limit $p\to 0$ (for a related study, see~\cite{Dudal:2012zx}); 
the result for  $A(-k,0,k)$ is shown in the left panel of~\fig{fig3}.

Even though the increase with respect to the tree level value is not too dramatic, the 
peak of $A(-k,0,k)$ at around 1 GeV turns out to be just right: 
when coupled with \1eq{tt2}, a considerable enhancement of the 
resulting $F(p^2)$ is produced, due to the nonlinear nature of this system of integral equations.
As shown in the right panel of~\fig{fig3}, one may actually reproduce the lattice data 
very accurately, using the correct value for the strong coupling $\alpha_s$~\cite{Boucaud:2005rm,Boucaud:2008gn}.

\section{\label{quark}Quark sector}

\bigskip
\indent

It is well known that the nonperturbative mechanism that endows quarks with their constituent masses accounts 
for about $98\%$ of the visible matter in the Universe. In fact, it has often been denominated as 
the most efficient mass generating mechanism known, practically generating masses from nothing
(see~\cite{Cloet:2013jya} and references therein).
This fundamental phenomenon (and the related dynamical breaking of the chiral symmetry) 
is encoded in the SDE of the quark propagator, know also as quark ``gap equation''. 

The standard decomposition of the quark propagator,    
$S(p)$, is given by~\cite{Roberts:1994dr}
\be
S^{-1}(p) =  A(p^2)\,\sla{p} - B(p^2) \mathbb{I} 
= A(p^2)[\sla{p}-{\mathcal{M}}(p^2) \mathbb{I}] \,,
\label{qpropAB}
\ee
where $\mathbb{I}$ is the identity matrix, and $\mathcal{M}(p^2)$ 
is the dynamical quark mass; its generation is inextricably connected with the  
dynamical breaking of the chiral symmetry.

\begin{figure}[!h]
\begin{center}
\includegraphics[scale=0.9]{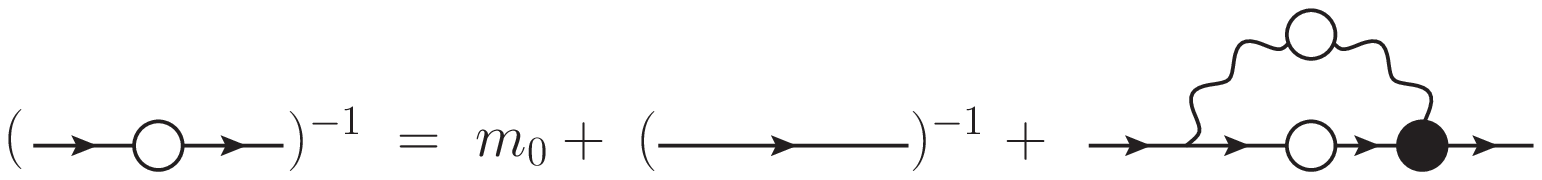}
\end{center}
\caption{\label{fig:quarksde} The SDE (gap equation) for the quark propagator, given by~\1eq{senergy}. The white blobs represent the
fully dressed gluon and quark propagators, while the black blob denotes the dressed quark-gluon
vertex.}
\end{figure}
%
 The SDE for the quark propagator is shown diagrammatically in~\fig{fig:quarksde}, and is given by
\begin{equation}
S^{-1}(p)= \sla{p} -m_0 -C_{\rm F}g^2\int_k
\Gamma_{\mu}^{[0]}S(k)\Gamma_{\nu}(-p,k,q)\Delta^{\mu\nu}(q) \,,
\label{senergy}
\end{equation}
where $q\equiv p-k$, $\Gamma_{\nu}(-p,k,q)$ is the fully dressed quark-gluon vertex,
$\Gamma_{\mu}^{[0]}$ is the tree-level value, and 
$C_{\rm F}$  is the Casimir eigenvalue in the fundamental representation.  
In this talk we will only consider the case of vanishing current quark mass, $m_0=0$, \ie  
the chiral symmetry is kept intact at the Lagrangian level. One of the characteristic features of this
integral equation is that, in order to give rise to nontrivial solutions,
the support of its kernel in a special momentum regime must overcome a critical amount.

As happens in the case of the ghost equation, the nonperturbative vertex constitutes a crucial ingredient 
for overcoming the aforementioned critical amount and arriving to a physically acceptable solution.
The situation, however, is much more complicated now, given that the tensorial decomposition of $\Gamma_{\nu}(-p,k,q)$ 
consists of twelve independent structures~\cite{Davydychev:2000rt} instead of only two. Projecting them out of the 
SDE that governs $\Gamma_{\nu}(-p,k,q)$  represents a major technical challenge, let alone 
solving the resulting intricate system of coupled integral equations. Lattice simulations, on the 
other had, have obtained some of the corresponding form factors for special 
kinematic configurations~\cite{Skullerud:2002ge,Skullerud:2003qu,Kizilersu:2006et},
but their relevance to the problem at hand is relatively limited.

\begin{figure}[t]
\begin{center}
\includegraphics[scale=0.8]{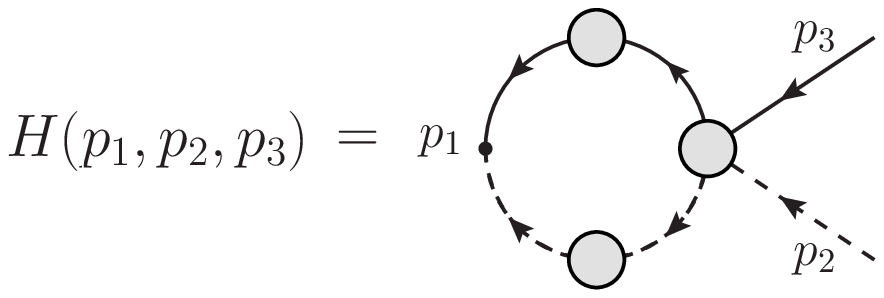}
\caption{Diagrammatic representation of the ghost-gluon kernel $H$.} 
\label{figH}
\end{center}
\end{figure}

The alternative that has been extensively adopted is to employ gauge-technique inspired Ans\"atze 
(see, \eg~\cite{Kizilersu:2009kg} and references therein)  
for $\Gamma_{\nu}(-p,k,q)$,
namely construct it in such a way that it satisfies the STI imposed by the BRST symmetry. In particular, 
$\Gamma_{\mu}(p_1,p_2,p_3)$  obeys the fundamental STI~\cite{Marciano:1977su}
\be
p_3^{\mu}\Gamma_{\mu}(p_1,p_2,p_3) = 
F(p_3)[S^{-1}(-p_1) H(p_1,p_2,p_3) - {\overline H}(p_2,p_1,p_3) S^{-1}(p_2)]\,,
\label{STI}
\ee
where the fermion-ghost scattering kernel  $H(p_1,p_2,p_3)$ is defined diagrammatically in ~\fig{figH},
and has the tensorial decomposition
\be
H(p_1,p_2,p_3) = X_0 \mathbb{I}  +X_1 \sla{p_1} +  X_2  \sla{p_2} + X_3 \tilde\sigma_{\mu\nu}p_1^{\mu} p_2^{\nu}, 
\ee
with $\tilde\sigma_{\mu\nu} \equiv \frac{1}{2}[\gamma_{\mu},\gamma_{\nu}]$;
${\overline H}(p_2,p_1,p_3)$ is the ``conjugated'' version of $H(p_1,p_2,p_3)$~\cite{Davydychev:2000rt}. 

Note that when both $F$ and $H$ are set to unity (\ie the ghost sector is switched off), \1eq{STI} reduces to 
the text-book WI relating the photon-electron vertex with the electron propagator in QED. The 
the so-called Ball-Chiu (BC) vertex~\cite{Ball:1980ay}, 
\bea 
\Gamma^{\mu}_{\rm BC}(p_1,p_2,p_3)&=&\frac{A(p_1)+A(p_2)}{2}\gamma^{\mu} \nonumber \\
&+&\frac{(p_1-p_2)^{\mu}}{p_1^2-p_2^2}\left\{\left[A(p_1)-A(p_2)\right] 
\frac{\sla{p_1}-\sla{p_2}}{2}
+\left[B(p_1)-B(p_2)\right] 
\right\} \,.
\label{bcvertex}
\eea
satisfies precisely this particular WI, instead of the full STI of \1eq{STI}, and has served as the 
starting point of numerous studies. 

Specifically, use of \1eq{bcvertex} into \1eq{senergy} leads to a coupled system for $A(p^2)$ and $B(p^2)$ 
\bea
A(p^2)&=& 1 + C_{F}g^2\,\int_{k}\,
\frac{{\cal K}(p-k)}{A^2(k^2)k^2+B^2(k^2)}{\cal K}_A^{\rm BC}(k,p)\,, 
\nonumber\\ 
B(p^2) &=& C_{F}g^2 \int_{k}\,\frac{{\cal K}(p-k)}{A^2(k^2)k^2+B^2(k^2)} {\cal K}_B^{\rm BC} (k,p)\,,
\label{scalar}
\eea
where the kernel ${\cal K}_B^{\rm BC} (k,p)$ originates from \1eq{bcvertex}
(for its explicit form, see, \eg~\cite{Aguilar:2010cn}), while  
the kernel ${\cal K}(q)$ encompasses all remaining contributions. 

\begin{figure}[h]
\begin{minipage}[b]{0.45\linewidth}
\centering
\includegraphics[scale=0.35]{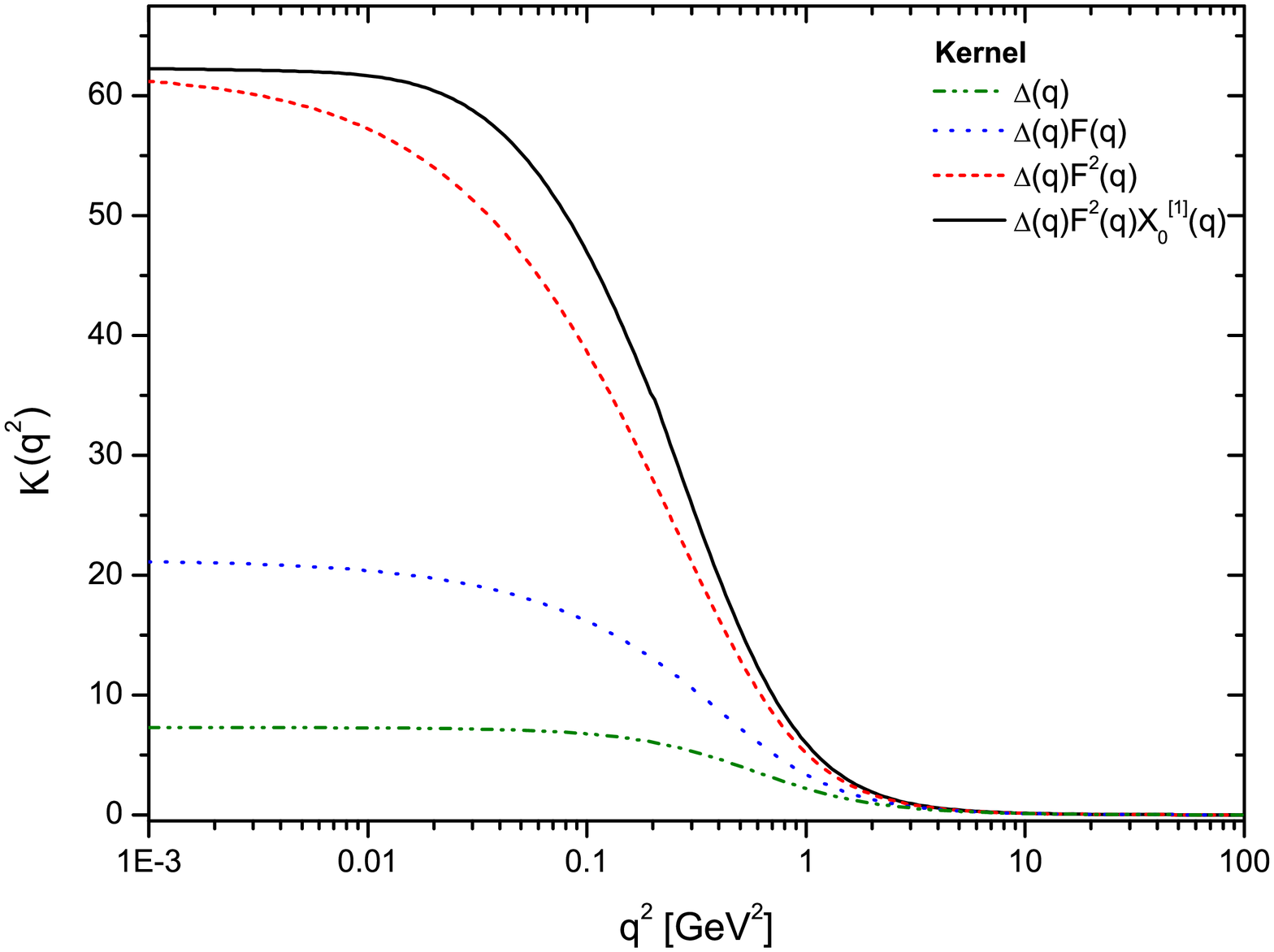}
\end{minipage}
\hspace{0.5cm}
\begin{minipage}[b]{0.50\linewidth}
\centering
\includegraphics[scale=0.35]{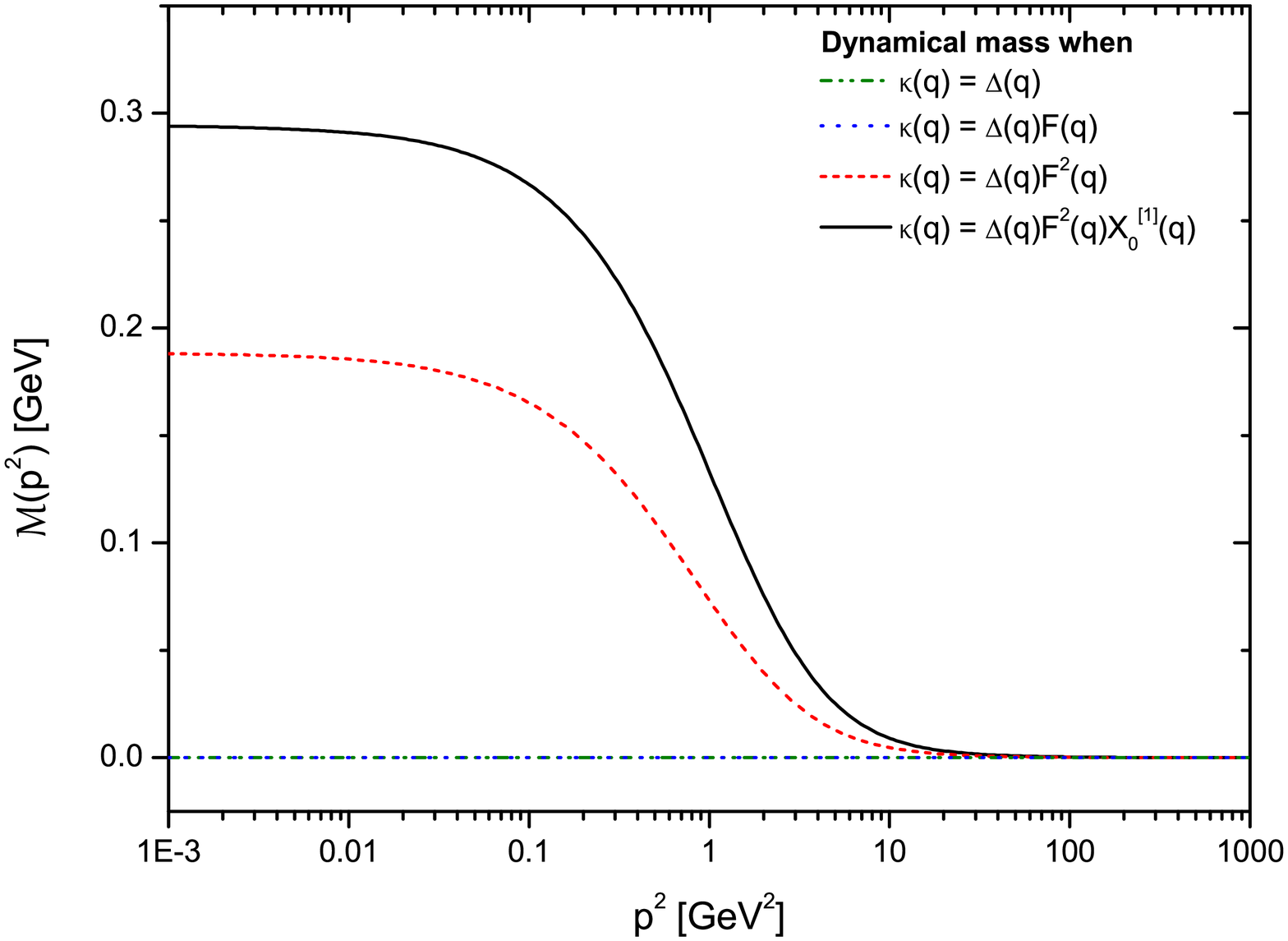}
\end{minipage}
\vspace{-0.5cm}
\caption{{\it Left panel}: The individual contribution of the 
ingredients composing $ {\cal K}(q)$. 
The green dotted-dashed line represents
the case where \mbox{${\cal K}(q)=\Delta(q)$}, the blue dotted line the 
 case where \mbox{${\cal K}(q)=\Delta(q)F(q)$}, in the red dashed curve 
\mbox{${\cal K}(q)=\Delta(q)F^2(q)$} and, finally the black 
continuous line represents the case where ${\cal K}(q)$ assumes 
the full form used, namely \mbox{${\cal K}(q)=\Delta(q)F^2(q)X_0^{[1]}(q)$}. 
{\it Right panel}: The corresponding 
dynamical quark mass generated when we use in \1eq{scalar} 
the different forms of ${\cal K}(q)$ shown in the left panel.}
\label{figqm}
\end{figure}

If the ghost contributions to $\Gamma_{\mu}$ are switched off, ${\cal K}(q)$ consists simply of 
$\Delta(q)$; in this case, however, the resulting combined kernel 
(right panel of~\fig{figqm}) fails to furnish a nontrivial solution (left panel of~\fig{figqm}).

Given this unsatisfactory situation, a next possible improvement consists in restoring 
the omitted ghost-related contributions to $\Gamma_{\mu}$, thus achieving the  
gradual ``non-Abelianization'' of the Ball-Chiu vertex. In particular,  
one may include the contributions of $F(q)$ and $H$, as dictated by the STI of \1eq{STI}. 
Evidently, since  $F(q)$ is very well known (see previous section), 
the main challenge is to obtain an approximation for $H$, or at least, for some of 
its form factors. In addition, renormalization introduces 
(effectively) into ${\cal K}(q)$ an 
additional multiplicative factor of  $F(q)$, an approximation 
which guarantees that the anomalous dimension of the resulting dynamical quark mass 
is consistent with the known one-loop expression (see also \cite{Fischer:2003rp}). On the right panel of~\fig{figqm} 
one may follow the gradual built-up of ${\cal K}(q)$ and on the left the  
form of the quark mass $\mathcal{M}(p^2)$ obtained at each step. Evidently, when 
${\cal K}(q) = F^2(q)\Delta(q)$, a considerable amount of $\mathcal{M}(p^2)$ is obtained from 
\1eq{scalar}; however, one deviates by a factor of about $30\%$  
from the phenomenologically accepted value.

\begin{figure}[t]
\begin{center}
\includegraphics[scale=0.4]{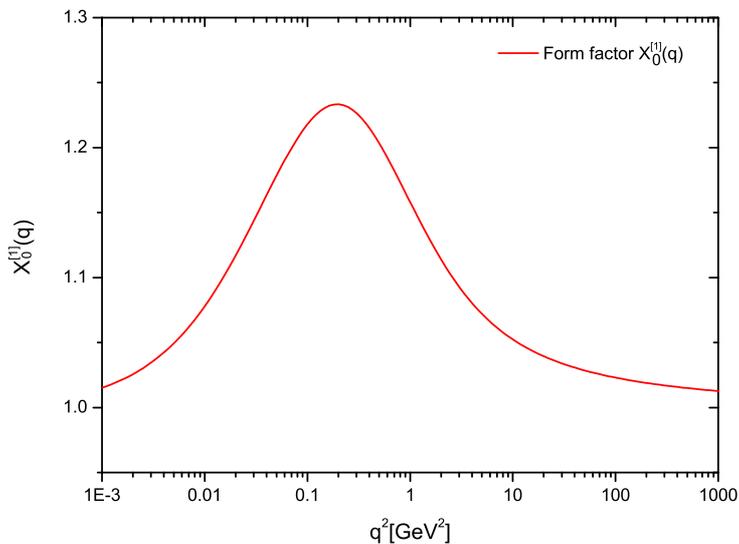}
\caption{Numerical result for the form factor $X_0^{[1]}(q)$, obtained 
within the one-loop dressed approximation, using fully dressed propagators and bare vertices~\cite{Aguilar:2010cn}.} 
\label{sk}
\end{center}
\end{figure}

It turns out that the missing amount of dynamical quark mass  
may be obtained by supplying to ${\cal K}(q)$ some of the strength coming from $H$. 
In particular, the form factor $X_0$ was 
computed within the one-loop dressed approximation, $X_0^{[1]}$, 
for the symmetric kinematic configuration $(-q/2,-q/2,q)$~\cite{Aguilar:2010cn}.
The result is shown in~\fig{sk}; note that
the deviation from unity (peaked around 600 MeV) is modest, but, due to the nonlinearity of the 
gap equation, it gives rise to a considerable amount of additional quark mass, 
as can be appreciated in~\fig{figqm}.

\section{\label{hadrons}Hadron observables from continuum QCD} 

\bigskip
\indent

\begin{figure}[!h]
\begin{center}
\includegraphics[scale=0.6]{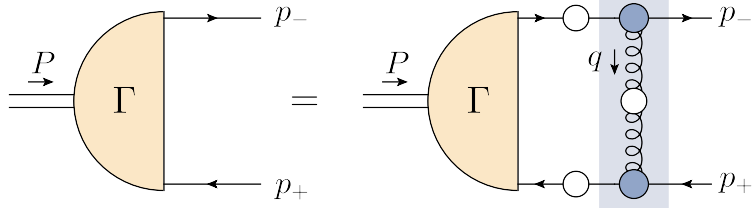}
\end{center}
\caption{\label{fig:BSkernel} A typical Bethe-Salpeter equation. The blue box contains the 
interaction kernel, whose form is crucial for the type of solutions (spectrum) obtained.}
\end{figure}

One of the longstanding challenges of QCD is to furnish quantitatively accurate {\it ab initio} predictions 
for the observable properties of hadrons~\cite{Chang:2009zb,Chang:2010hb,Qin:2011xq,Chang:2011ei}. In a recent work~\cite{Binosi:2014aea}, 
considerable progress has been made in this direction; in what follows we will highlight the main aspects 
of these new developments.

Let us consider the kernel appearing in 
a typical Bethe-Salpeter equation~\cite{Eichmann:2009zx}, shown in~\fig{fig:BSkernel}, 
which is contained in the blue box.
This kernel receives a ``universal'' (process-independent) contribution, 
whose origin is the pure gauge sector of the theory. In that sense, this contribution 
constitutes the common ingredient of any such kernel, regardless of the nature of the particles 
between which it is embedded; most notably, it does not depend on the valence-quark content of the Bethe-Salpeter equation. 

The systematic diagrammatic identification of the precise pieces that constitute this particular 
quantity may be carried out following the PT. As has been explained in detail in the literature, 
the upshot of this construction is the rearrangement of a physical amplitude 
into sub-amplitudes with very special properties~\cite{Binosi:2009qm}.
In addition, it is well-known 
that the result of this particular rearrangement coincides 
with the $BB$ gluon propagator defined in the BFM (see~\cite{Binosi:2002ft} and references therein).

Specifically, the standard $\Delta(q^2)$ introduced in~\1eq{prop} and the 
corresponding quantity of the $BB$ gluon propagator, denoted by $\widehat\Delta (q^2)$,
are related by the exact relation 
\be
\Delta(q^2) = \widehat\Delta(q^2) [1+G(q^2)]^2\,,
\label{BQI}
\ee
which is completely analogous to~\1eq{BQIs}.

At the one-loop level, and keeping only ultraviolet logarithms, one has~\cite{Aguilar:2008xm}
\bea
1+G(q^2) &=& 1 +\left(\frac{9}{4}\right)
\frac{\alpha_s C_{\rm {A}}}{12\pi} \ln\left(\frac{q^2}{\mu^2}\right),
\nonumber \\
\Delta^{-1}(q^2) &=& q^2\left[1+ \left(\frac{13}{2}\right)
\frac{\alpha_s C_{\rm {A}} }{12\pi}\ln\left(\frac{q^2}{\mu^2}\right)\right]^{-1}, 
\label{pert_gluon}
\eea 
and thus
\be
\widehat\Delta^{-1}(q^2) = q^2 \left[1+ b \alpha_s\ \ln\left(\frac{q^2}{\mu^2}\right)\right].
\ee
where $b = 11 C_A/12\pi$ is the first coefficient of the Yang-Mills  $\beta$ function, as it should~\cite{Abbott:1980hw}.

Due to the special WIs satisfied by the PT-BFM Green's functions, 
the (dimensionful) universal combination
\be
{\widehat d}(q^2) = \alpha_s \widehat\Delta(q^2)
= \frac{\alpha_s \Delta(q^2)}{\left[1+G(q^2)\right]^2}
\label{defd}
\ee
is renormalization-group invariant.
In particular, the renormalization constants of the gauge-coupling and of $\widehat\Delta^{-1}(p^2)$, 
defined as 
\bea
g(\mu^2) &=& Z_g^{-1}(\mu^2) g_0 ,\nonumber \\
\widehat\Delta(p^2,\mu^2) & = & \widehat{Z}^{-1}_A(\mu^2)\widehat{\Delta}_0(p^2), 
\label{conrendef}
\eea
where the ``0'' subscript indicates bare quantities, satisfy the 
QED-like relation 
\be
{Z}_{g} = {\widehat Z}^{-1/2}_{A}.
\label{ptwi}
\ee
Therefore, 
\be
{\widehat d}_0(q^2) = {\alpha_s}_0 \widehat\Delta_0(q^2) = \alpha_s(\mu^2) \widehat\Delta(q^2,\mu^2) = {\widehat d}(q^2)
\label{ptrgi}
\ee
maintains the same form before and after renormalization, {\it i.e.}, it 
forms a $\mu$-independent quantity.

\begin{figure}[t]
\begin{minipage}[b]{0.45\linewidth}
\centering
\includegraphics[scale=0.35]{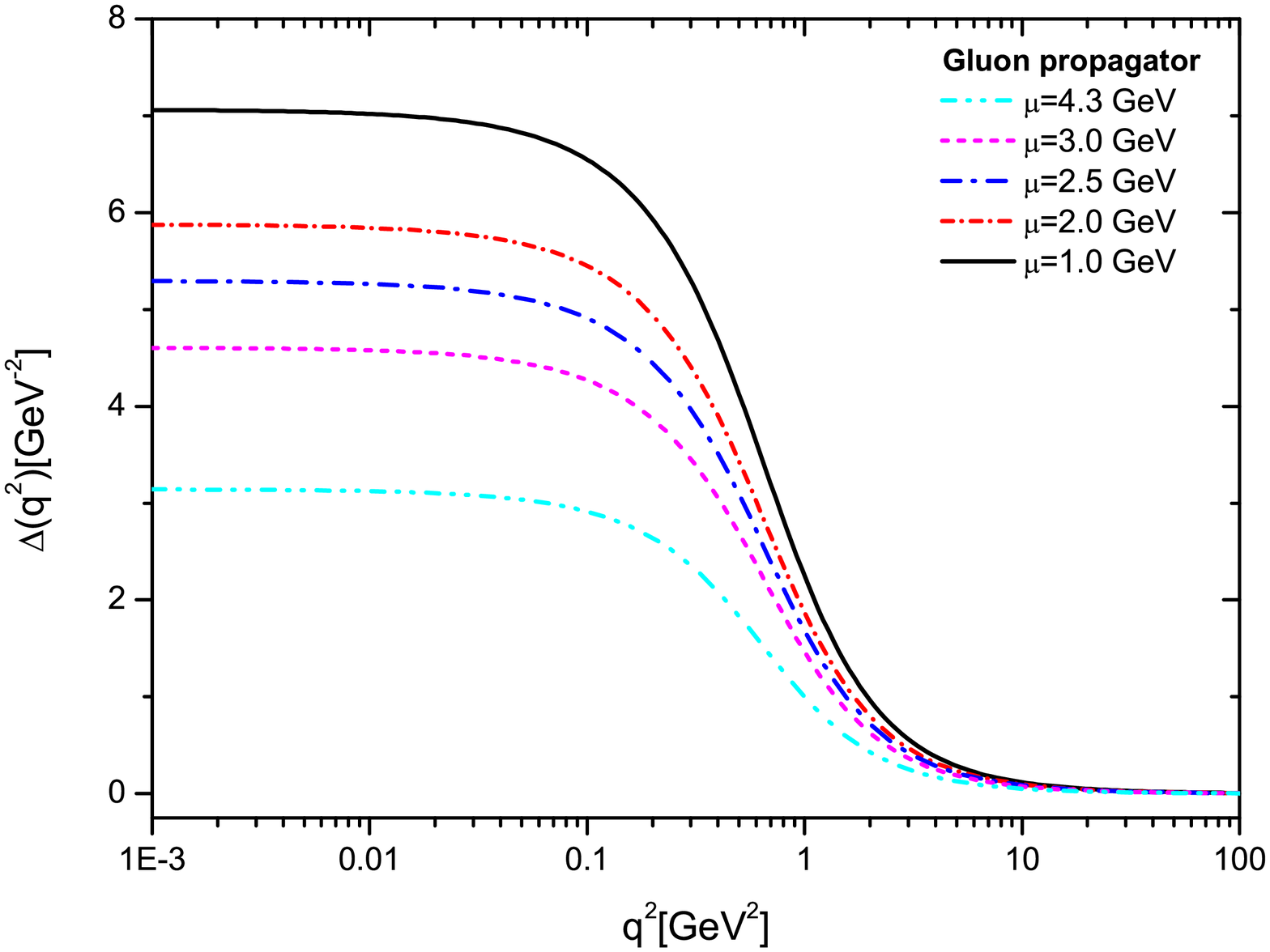}
\end{minipage}
\hspace{0.5cm}
\begin{minipage}[b]{0.50\linewidth}
\includegraphics[scale=0.35]{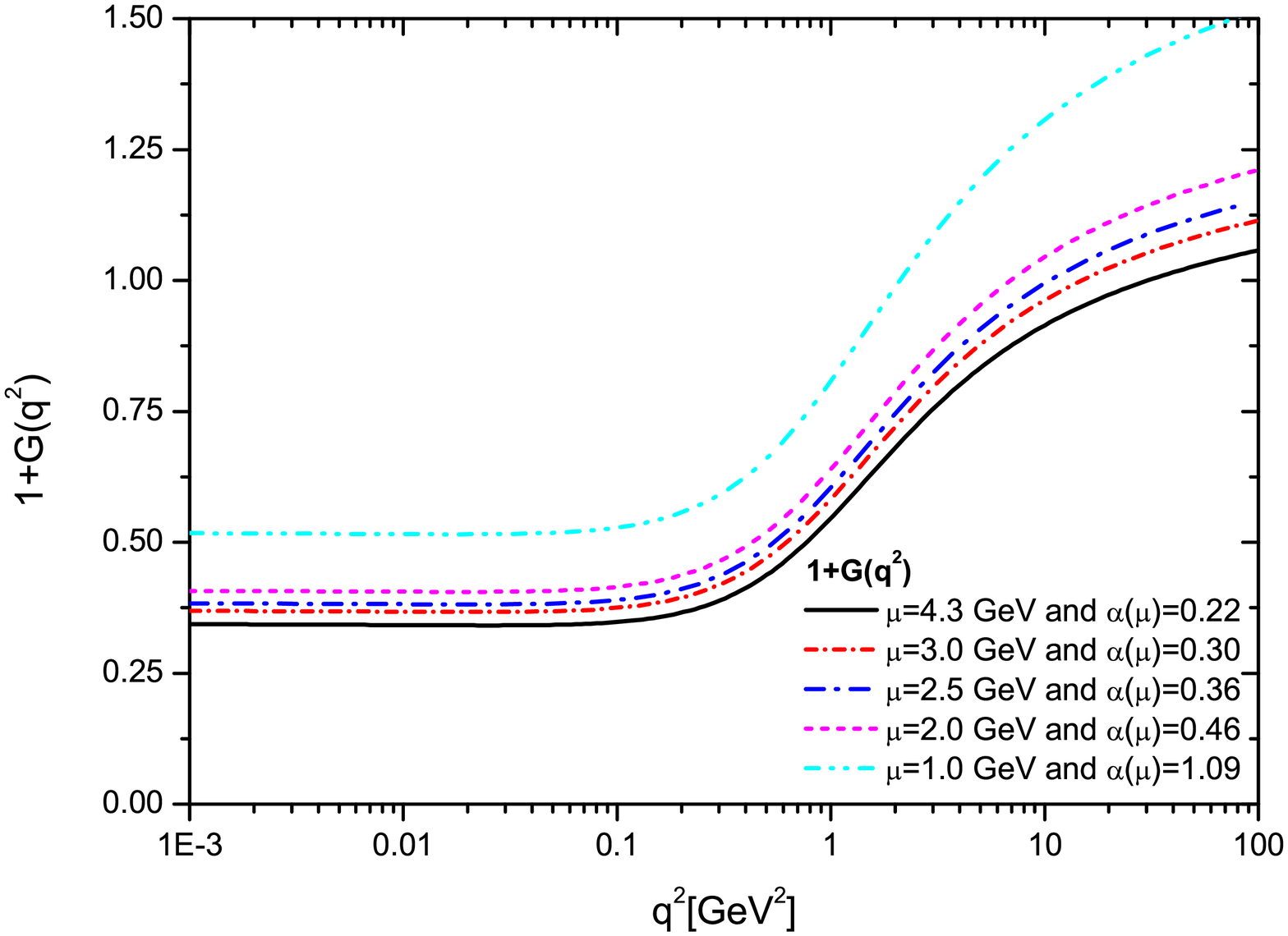}
\end{minipage}
\vspace{-0.25cm}
\caption{\label{fig:rgiing} {\it Left panel}:: The gluon propagator for different values of the renormalization point $\mu$. 
{\it Right panel}: The auxiliary function $1+G(q^2)$ for the same set of values of $\mu$.}
\end{figure}

\begin{figure}[t]
\begin{center}
\includegraphics[scale=0.4]{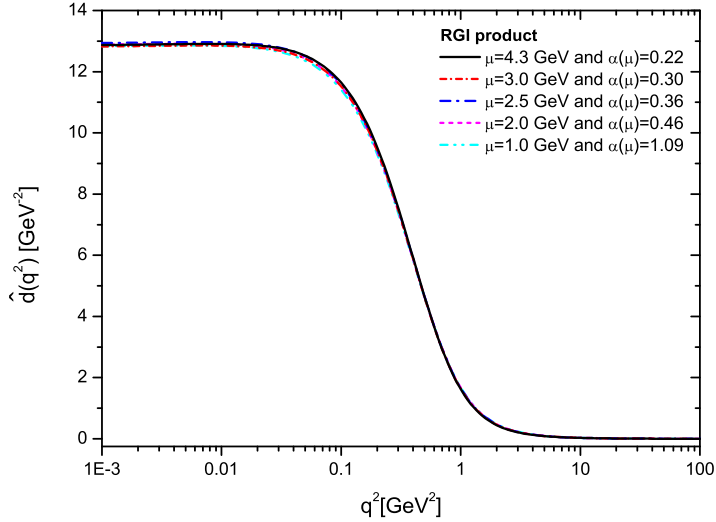}
\end{center}
\caption{\label{fig:rgi} The RG-invariant product ${\widehat d}(q^2)$ 
put together according \1eq{defd}, using the results of \fig{fig:rgiing}:
one practically obtains one single curve for all  different values of the renormalization point $\mu$}
\end{figure}

This last property has been explicitly verified to an impressive accuracy, 
following the procedure summarized in~\fig{fig:rgiing} and~\fig{fig:rgi}.
The gluon propagator $\Delta(q^2)$, obtained from the lattice data of~\cite{Bogolubsky:2007ud}, is 
multiplicatively renormalized (in the momentum-subtraction scheme) 
at five different values of $\mu$. The dynamical equation furnishing $1+G(q^2)$ 
(not reported here, see~\cite{Aguilar:2009pp,Aguilar:2009nf}) is subsequently solved, for the same values of  $\mu$.
When the resulting curves (given in the two panels of~\fig{fig:rgiing}) are combined 
according to~\1eq{defd}, it turns out that a unique curve emerges, provided that the 
values of each $\alpha_s(\mu^2)$  are those shown in the insert of ~\fig{fig:rgi}.


From the dimensionful quantity ${\widehat d}(q^2)$ one may extract a dimensionless 
effective charge, to be denoted by ${\cal I}(q^2)$, by simply pulling out a factor of $q^2$, namely
\be
{\cal I}(q^2) = q^2{\widehat d}(q^2)\,.
\label{effch}
\ee

This SDE-derived quantity allows for a direct comparison with an analogous quantity commonly employed by practitioners for describing 
the (momentum-dependent) interaction between quarks, obtained
through a well-defined truncation of the equations in the matter sector that are relevant to bound-state properties.
In particular, one uses 
\begin{equation}
{\cal I}(q^2)=k^2 {\cal G}(q^2);\quad
{\cal G}(q^2)=\frac{8\pi^2}{\omega^4}D{\mathrm e}^{-q^2/\omega^2}+\frac{8\pi^2\gamma_m (1-{\mathrm e}^{-q^2/4m_t^2})}{q^2\ln[\tau+(1+q^2/\Lambda_\s{\rm QCD}^2)^2]},
\label{bu}
\end{equation}
where $\gamma_m=12/(33-2N_f)$ [typically, $N_f=4$], $\Lambda_\s{\rm QCD}=0.234$ GeV; \mbox{$\tau={\rm e}^2-1$}, $m_t=0.5$ GeV. A large body of observable properties of ground-state vector- and isospin-nonzero pseudoscalar mesons (and even numerous properties of the nucleon and $\Delta$ resonance) are practically insensitive to variations of $\omega\in[0.4,0.6]$ GeV, so long as $\varsigma^3=D\omega=$const. The value of $\varsigma$ is typically chosen to obtain the measured value of $f_\pi$; however its value depend on the truncation scheme used when deriving~(\ref{bu}). The usual rainbow-ladder (RL) truncation~\cite{Maris:1999nt}
yields $\varsigma_\s{\rm RL}=0.87$ GeV, while the improved truncation scheme (DB) of~\cite{Chang:2009zb} gives $\varsigma_\s{\rm DB}=0.55$ GeV.

\begin{figure}[t]
\centerline{\includegraphics[width=0.65\linewidth]{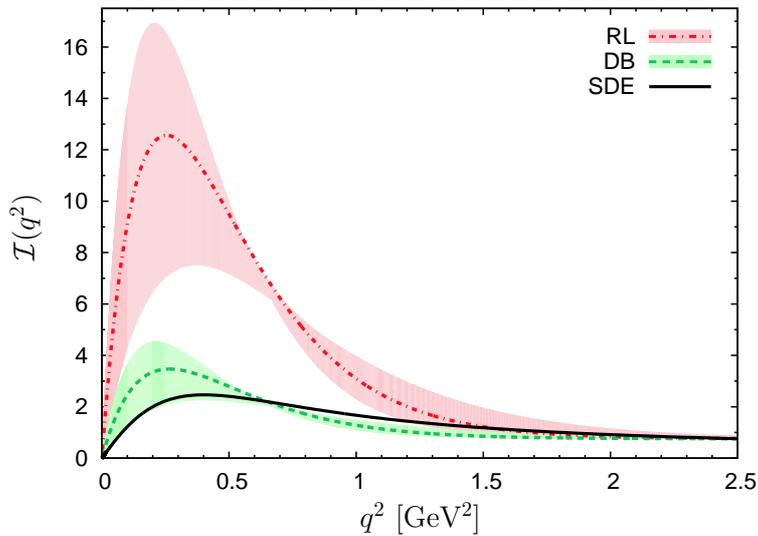}}
\caption{\label{figRL}Comparison between the SDE-derived effective interaction obtained from  
the gauge-sector~[Eq.~(\ref{effch}), solid, black] with those obtained using Eq.~(\ref{bu}): \emph{pale-red band}: RL truncation, 
\emph{pale-green band}: DB truncation.  
The bands denote the domain of constant ground-state physics $0.45<\omega<0.6$, with the curve within each band obtained with $\omega=0.5\,$GeV.}
\mbox{}\vspace{-.8cm}
\end{figure}

As can be seen in~\fig{figRL}, 
the direct comparison between the ${\cal I}(q^2)$ defined through~\1eq{effch} and~\1eq{bu} reveals an excellent agreement.
This seems to suggest that a QCD-derived description of measurable hadron properties is well within our reach.

\section{\label{concl}Conclusions}

\bigskip
\indent

In this presentation we have reviewed a large body of developments related to  
the study of the off-shell Green's functions of pure Yang-Mills theories 
and QCD, using SDEs and the lattice simulations in the Landau gauge.
The  general picture that surfaces may be summarized as follows: 

{\it i)} Gluons just as quarks acquire a dynamical mass, 
which is controlled by the corresponding ``gap equation''. 

{\it ii)}
Ghosts remain massless, and have a finite dressing function that 
can be accurately obtained from the corresponding SDE.
 
{\it iii)}
The quality of the ingredients entering into the quark gap equation is steadily improving,
giving rise to phenomenologically successful quark masses.  
 
{\it iv)}
A significant step has been taken for bridging
the gap between nonperturbative continuum QCD predictions and hadron phenomenology.

Even though a lot of work remains to be done in all directions, 
Richard Feynman's famous quote  
from the first of his seven ``Messenger Lectures'' at Cornell University in 1964 comes to mind: {\it Nature uses only the longest threads 
to weave her patterns, so that every small piece of the fabric reveals the organization of the entire tapestry}''.  
We seem to have reached a similar point with QCD: even though the Green's functions 
that have been studied are but a small piece of the QCD fabric, unequivocal signs of an 
underlying organization begin to emerge.

\ack 

\bigskip
\indent

This research is supported by the Spanish MEYC under 
grant FPA2011-23596 and the Generalitat Valenciana under grant “PrometeoII/2014/066”.
I would like to thank the organizers of the ``Discrete 2014'' for their kind invitation and hospitality,
and all participants for providing a most stimulating and pleasant atmosphere.

\end{document}